\documentclass[aps,pre,floatfix,longbibliography]{revtex4-2}
\usepackage{amsfonts}
\usepackage{amsmath}
\usepackage{amsthm}
\usepackage{xcolor}
\usepackage{graphics}
\usepackage{graphicx,epstopdf}
\usepackage[export]{adjustbox}
\usepackage{subfigure}
\usepackage{amssymb}
\usepackage[colorinlistoftodos]{todonotes}  
\usepackage[T1]{fontenc}
\usepackage{amsmath,amssymb,epsfig,latexsym,graphicx,dcolumn}
\usepackage{natbib}
\usepackage{todonotes}
\usepackage{svg}
\usepackage{printlen}
\usepackage{footmisc}
\usepackage{hyperref}
 
\begin{document}
\setcounter{page}{1}
\title{Direct numerical simulation benchmarks for the prediction of \\ boundary layer bypass transition in the narrow sense }

\author{Xiaohua Wu} 
\affiliation{Royal Military College of Canada, }
\author{Carlos A. Gonzalez} 
\affiliation{Center for Turbulence Research, Stanford University,} 
\author{Rahul Agrawal}
\email{Email: rahul29@stanford.edu}
\affiliation{Center for Turbulence Research, Stanford University }


\date{\today}

\begin{abstract}
We report a comprehensive set of direct numerical simulation benchmarks of bypass transition in the narrow sense with inlet freestream turbulent intensity levels $0.75\%$, $1.5\%$,  $2.25\%$,  $3.0\%$ and $6.0\%$, respectively. Detailed descriptions of length scales and the rate of viscous dissipation are provided. We ask two key physical questions. First, how do the decay rates and length scales of freestream turbulence over a transitional and turbulent boundary layer compare to those in spatially developing isotropic turbulence without the wall? Second, what bypass mechanisms drive turbulent spot inception at the intermediate rage of freestream turbulence intensity level? We find that the boundary-layer freestream turbulence decay and length scales evolve similarly to their spatially developing isotropic turbulence flow without the wall counterparts. We also present evidence of the coexistence of two turbulent spot inception mechanisms at the inlet FST level $2.25\%$: the long low-speed streak primary and secondary instabilities (only in lower inlet FST levels) and the self-amplifying process of oblique vortex filaments interacting with a Delta-shaped low-speed patch underneath (prevailing only in higher inlet FST levels).  \\

 \end{abstract}
\maketitle

\section{Introduction}
  
Predicting boundary-layer transition is essential in several industry applications, including over aircraft wings, compressors, and turbine blades. To this end, transition modeling remains an active area of research. A crucial step in developing such models is their evaluation over zero-pressure-gradient, smooth-walled, flat-plate boundary layer (ZPGSFPBL) under isotropic freestream turbulence (FST) with varying inlet intensity levels ($\mathrm{FSTI}_{in}$).  \citet{wu2023new} classified such flows as boundary-layer bypass transition in the narrow sense to distinguish them from bypass transition arising from other phenomena, including laminar separation bubbles, roughness elements, or very high-level FST. \\ 


\noindent
Generally, the experiments of \citet{roach1990influence} are used as a benchmark for calibrating and evaluating transition prediction models (see \citet{westin1997application}, \citet{suzen2000modeling}, \citet{menter2006transition}, \citet{durbin2012intermittency}, \citet{ge2014bypass}, \citet{menter2015one}). \citet{roach1990influence} reported wind-tunnel experiments in three conditions: case T3A with upstream turbulence intensity $3.5\%$, case T3B with upstream turbulence intensity $6.5\%$, and case T3A- with upstream turbulence intensity $0.8\%$. Interestingly, the skin friction coefficient, $C_f$, in these experiments collapses onto the Blasius laminar flow solution before breakdown. However, despite its varied usage, several notable drawbacks are still associated with this dataset. For instance, the $C_f$ was estimated indirectly from momentum balance rather than from a measurement of the near-wall velocity gradient. Further, there are only a few recorded data points within the transition zone  ($C_f$ departs from the laminar solution and begins approaching the turbulent level), which is perhaps concerning as there is a wide gap in the freestream turbulence levels between cases T3A and T3B. Thus, there has been a need for additional reference datasets for use in transition model benchmarking and calibration. \\

\begin{figure}
    \centering\includegraphics[width=0.7\linewidth]{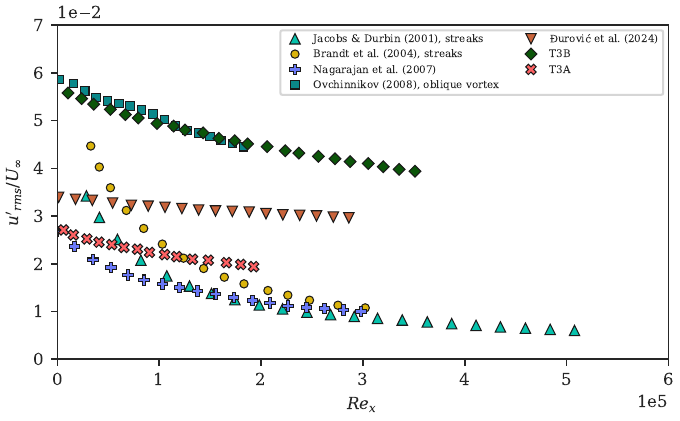}
    \caption{Decay of freestream turbulence intensity in the previous studies on bypass transition in the narrow sense. }
    \label{fig:fst_decay_reference}
\end{figure}

\noindent
In this spirit, from a computational standpoint, many direct numerical simulations (DNS) have been previously performed to study the transitioning boundary layer. We highlight the potential imperfections of such previous studies (see Fig.~\ref{fig:fst_decay_reference}). For instance, in the previous DNSs of \citet{jacobs2001simulations,nagarajan2007leading,brandt2004transition}, the freestream turbulence decays much faster than the experiments \citep{roach1990influence}. A consequence of the overly rapid decay of freestream turbulence is that the downstream stations (where the flow transitions) may experience lower turbulence intensities than desired. We briefly remark that previously, \citet{ovchinnikov2008numerical}  performed DNS of the transitioning boundary layer while including a leading edge and nearly reproduced the decay rates of T3B experimental case \citep{roach1990influence} for $\mathrm{FSTI}_{in} \sim 6 \%$. \citet{ovchinnikov2008numerical} perform two simulations, one with a ``full'' domain, and one with a ``symmetry plane'' (about the midplane of the leading edge). In the laminar flow region, their $C_f$ compares well with the Blasius solution in the full domain, but when the flow transitions, the $C_f$ does not agree with the experiments \citep{roach1990influence}. In the half-domain case, the simulations don't collapse onto the Blasius $C_f$ in the laminar region. The spatial decay rate of the freestream turbulence in the DNS of \citet{djurovic2024direct} also compares reasonably with \citet{roach1990influence}. However, their $C_f$ also does not collapse onto the Blasius solution before breakdown. \\

\noindent
The present study contributes toward addressing these limitations. To the authors’ knowledge, the database herein covers a wider than currently published range of $\mathrm{FSTI}_{in}$ conditions simulated under one computational framework. Further, we seek to answer two key physical questions: (i) how do the decay rates and length scales of freestream turbulence developing over a boundary layer compare to those in spatially developing isotropic turbulence? i.e., does the growth of the boundary layer affect the freestream turbulence? and (ii) what mechanisms drive turbulent spot inception at an intermediate range of $\mathrm{FSTI}_{in}$? \\

\noindent
Fundamentally, it is conjectured that the unsteady streamwise streaks dominate the boundary-layer bypass transition in the narrow sense, as supported by smoke flow visualizations, measurements of turbulence intensity profiles, and the non-modal theory on disturbance energy growth (the reader is referred to the works of \citet{kendall1998experiments,westin1994experiments,matsubara2001disturbance,fransson2020effect}). Although smoke visualizations often reveal the existence of long streaks, and measurements of turbulence intensities and correlations often show agreement with algebraic growth theory, the intricate processes of Blasius boundary-layer breakdown and the inception of the one infant turbulent spot at the smallest possible streamwise Reynolds number for a given instant/cycle are difficult to capture in experiments, nor can these processes be definitively predicted by existing theory. \\

\noindent
\citet{wu1999simulation} observed turbulent spots in DNS of ZPGSFPBL bypass transition and tracked the inception process of the spots. Following this work, \citet{djurovic2024direct,jacobs2001simulations,brandt2004transition,nagarajan2007leading,ovchinnikov2008numerical,wu2017transitional,wu2023new,alarcon2024role} have studied the flow over a transitioning boundary layer. 
\citet{jacobs2001simulations} attempted to reproduce the T3A experiment with an FSTI of $3.5\%$, and they found that infant spots arise from localized instabilities of low-speed streaks when the streaks interact with freestream turbulence. \citet{brandt2004transition} considered FSTI values of $1.5\%$, $3\%$, and $4.7\%$, and reported that Blasius layer breakdown is related to local secondary instabilities of long low-speed streaks in the form of either the sinuous-type driven by the spanwise shear or the varicose-type by the wall-normal shear. \citet{nagarajan2007leading} used FSTI values $3.5\%$ and $4.5\%$, and included the leading edge in their compressible flow simulations. They found that at a low level of FSTI, breakdown follows the streak-centered findings of \citet{jacobs2001simulations}  and \citet{brandt2004transition}. However, at a high level of FSTI, breakdown does not involve low-speed streaks. \citet{ovchinnikov2008numerical} studied the $\mathrm{FSTI}_{in}=6.7\%$ case, similar to the T3B experiment to find that although low-speed streaks exist, infant turbulent spots are detected upstream of the region of the streaks, and that spot precursors are traced to spanwise structures. These structures reorient to become hairpin vortices, which break down into infant spots. On the contrary, \citet{alarcon2024role} recently reported the convective evolution of the secondary instabilities of low-speed streaks using their DNS at $\mathrm{FSTI}_{in} = 3.45\%$, also including a leading edge. They presented statistics supporting the dominance of streak secondary instabilities in the breakdown of the numerous low-speed streaks. However, their $C_{f}$ notably differs from the Blasius solution in the early transition region. \\ 

\noindent   
\citet{wu2017transitional} simulated the $\mathrm{FSTI}_{in}= 3.0\%$ flow to find that the laminar boundary-layer breakdown is driven by a self-amplifying process of oblique vortex filaments interacting with a $\Delta$-shaped low-speed patch underneath. The oblique vortex filaments ($\Lambda$ vortex) flank the $\Delta$ patch. This process does not involve long, low-speed streaks. Sparse and chaotic low-speed streaks develop downstream of infant spots. These are consistent with the findings of \citet{ovchinnikov2008numerical} at FSTI $6.7\%$. Further, \citet{wu2023new} found that at $\mathrm{FSTI}_{in} = 1.5\%$, the laminar boundary layer breaks down when asymmetric inclined boundary-layer vortex filaments appear on one side of long low-speed streaks or when symmetric boundary-layer vortex filaments partially wrap the tail portion of a long low-speed streak (consistent with \citet{brandt2004transition}). The wavy appearance of streaks associated with the asymmetric inclined vortex filaments and symmetric vortex filaments are in line with the notions of ‘sinuous mode’ and ‘varicose mode’ of \citet{swearingen1987growth}, respectively. \citet{wu2023new} suggested that the streak-instability spot inception mechanism may be relevant for $\mathrm{FSTI}_{in}$ levels ranging from $0.5\%$ to $2\%$, and the mechanism of self-amplifying oblique-vortex interaction with an underneath $\Delta$-shaped low-speed patch may be more relevant for $\mathrm{FSTI}_{in}$ levels ranging from $2.5\%$ to $5\%$. \\

\noindent
Given these findings, we seek to ask how the switch from the non-streak-dominated mechanism to the streak-dominated mechanism occurs with the reduction of $\mathrm{FSTI}_{in}$. Alternatively, at some intermediate $\mathrm{FSTI}_{in}$, would one breakdown mechanism dominate the other, or would the two mechanisms coexist in the same flow but at different instants/cycles? In this spirit, we also seek to answer if the developing boundary layer affects the freestream above it, which may, in turn, affect the transition process. \\

\noindent
The rest of this work is organized as follows: Section II describes the computational method. Section III briefly discusses statistics of the DNS database including skin friction, and length, dissipation scales. Section IV assesses the decay of spatially evolving freestream turbulence with and without walls. Section V discusses the turbulent spot inception mechanisms, including at an intermediate level ($\approx 2.25\%$) of $\mathrm{FSTI}_{in}$. Finally, concluding remarks are provided in Section VI. 

\section{Computational Method}

\begin{figure}
     \centering
     \includegraphics[width=0.5\linewidth]{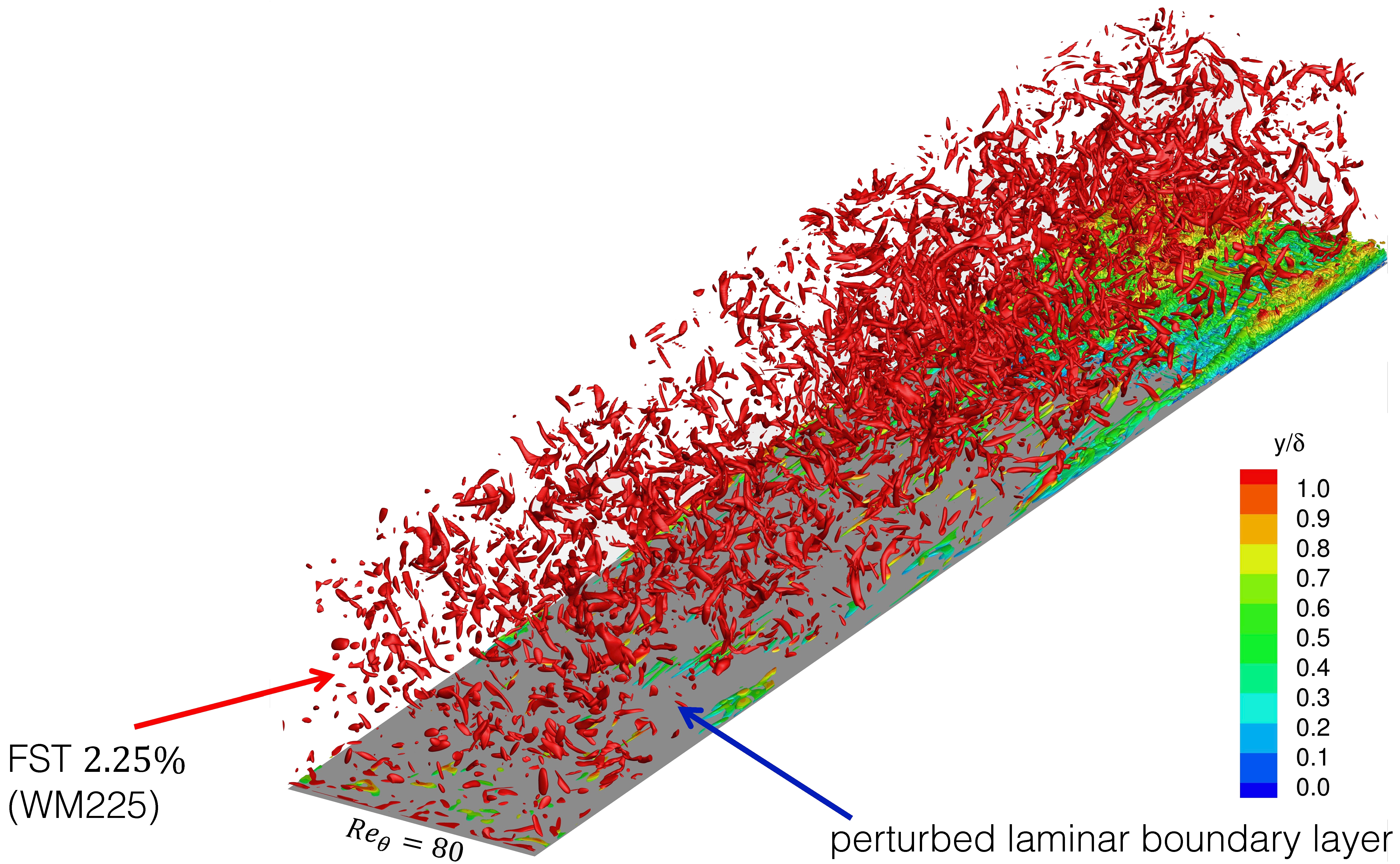}
     \caption{Isosurfaces of swirling strength, as defined in \citet{adrian2007hairpin}, in WM225 over the transitional region. }
     \label{fig:view_dns_3d_WM225}
 \end{figure}

\begin{figure}
    \centering
    \includegraphics[width=0.7\linewidth]{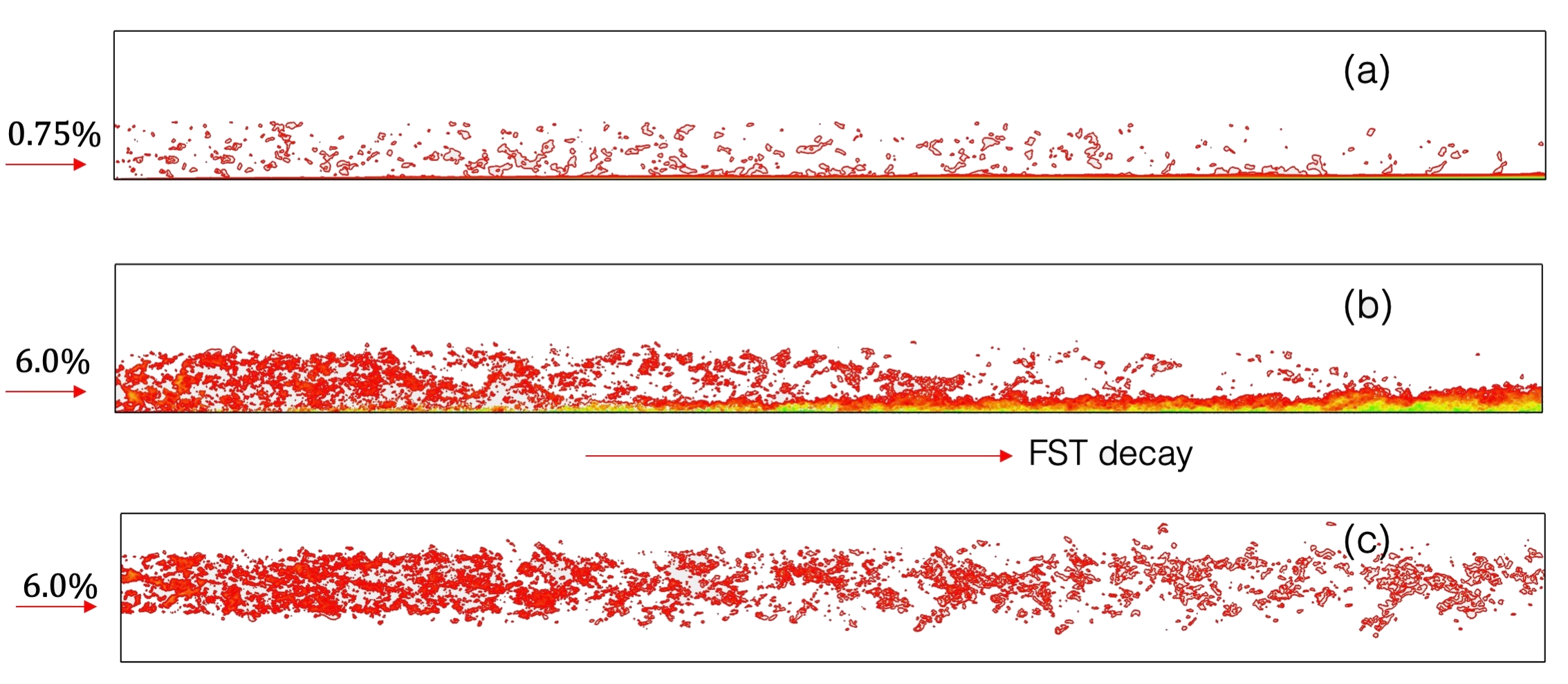}
    \caption{Visualization of three DNS cases using contours of streamwise velocity $u$ over a random $xy$-plane where blue and red colors indicate $u=0$ and 
    $u=U_{\infty}$, respectively. (a) WM075; (b) WM600; (c) SDIT600. }
    \label{fig:wm075_wm600_sdit600_u2d}
\end{figure}

We perform five DNSs of boundary-layer bypass transition in the narrow sense and three additional DNSs on stand-alone spatially developing isotropic turbulent flow without a wall. The $\mathrm{FSTI}_{in}$ levels for the five cases are $0.75\%$ (denoted as WM075), $1.5\%$ (denoted as WM150), $2.25\%$ (denoted as WM225), $3.0\%$ (denoted as WM300) and $6.0\%$ (denoted as WM600), respectively (see Fig.~\ref{fig:view_dns_3d_WM225} and Figs.~\ref{fig:wm075_wm600_sdit600_u2d}(a) \& (b)). The $\mathrm{FSTI}_{in}$ levels of spatially-developing isotropic turbulence without the wall are $1.5\%$ (denoted as SDIT150), $3.0\%$ (denoted as SDIT300), and $6.0\%$ (denoted as SDIT600), respectively (also see Fig.~\ref{fig:wm075_wm600_sdit600_u2d}(c)).\\

The domain size, grid resolutions, inflow turbulence generation, boundary conditions, and numerical method for the WM-series DNSs are identical to those in \citet{wu2017transitional,gonzalez2024benchmarks} and skipped for brevity (as a marker for the sufficiency of the resolution, we briefly highlight that the spatial resolutions are smaller than $4\eta$ in wall parallel and wall-normal directions, including in the freestream for WM150 flow. Further, the temporal resolution is less than $\tau_{\eta}=(\nu/\varepsilon)^{1/2}$, or the local Kolmogorov time scale.)  \\

\begin{figure}
    \centering
    \includegraphics[width=1.0\linewidth]{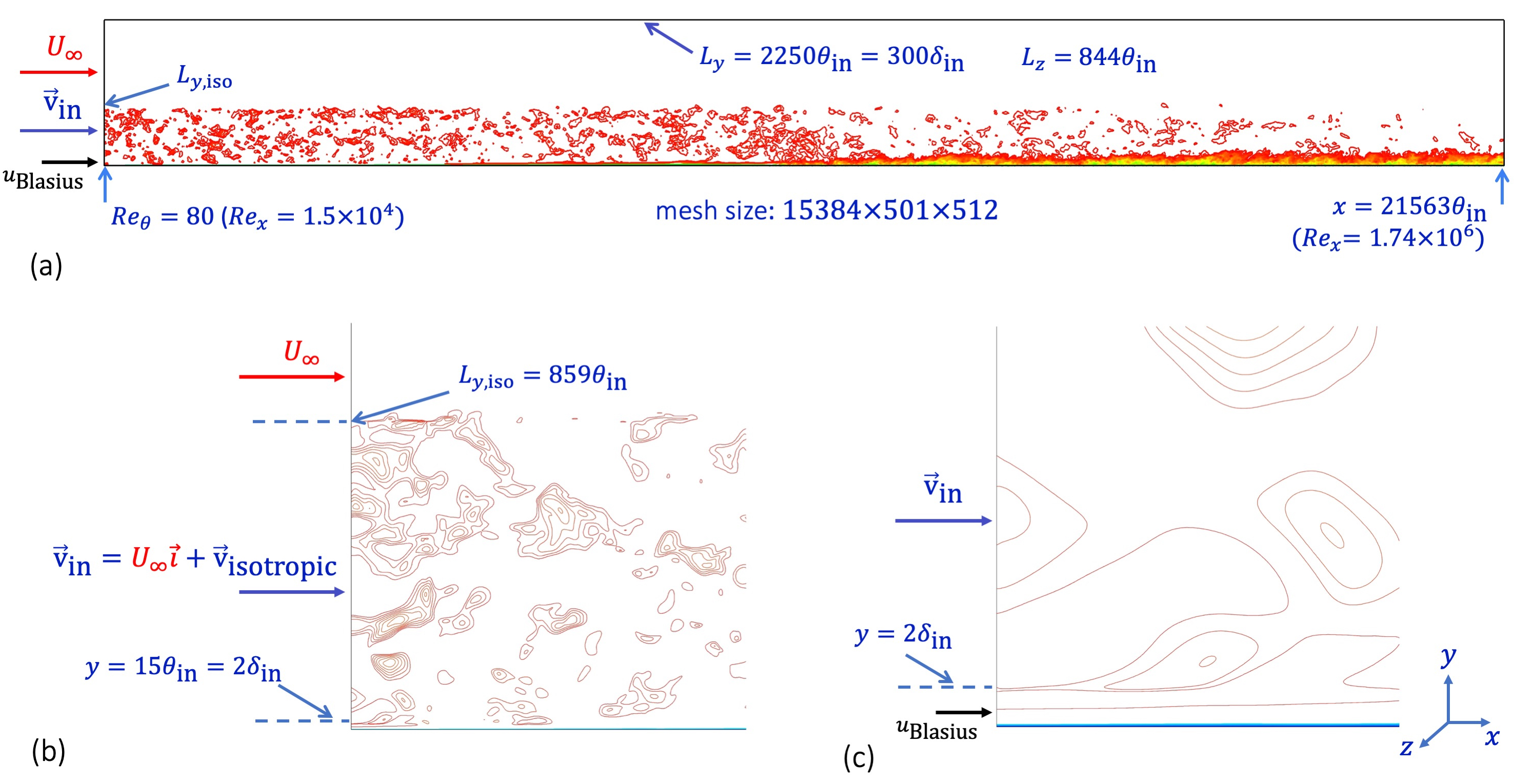}
    \caption{Illustration of boundary layer DNS design using contours of streamwise velocity $u$ over a random $xy$-plane where blue and red colors indicate $u=0$ and 
    $u=U_{\infty}$, respectively. (a) full view; (b) zoomed view near the inlet; (c) further zoomed view near the inlet. }
    \label{fig:inlet_design_dns}
\end{figure}

\noindent
Fig.~\ref{fig:inlet_design_dns} recapitulates some important parameters for the WM-series DNS. FST is introduced at the inlet over the wall-normal range $15\theta_{in} <y<L_{y,iso}$, where $\theta_{in}$ is the constant inlet boundary-layer momentum-thickness. Below and above this height, a uniform inflow ($U_{\infty}$, without fluctuations) and the mean Blasius velocity profile are fed, respectively. At the streamwise exit, domain height $L_{y}$ is equivalent to $24.31$, $9.05$, $7.36$, $7.24$, and $6.66$ local boundary-layer thickness $\delta_{exit}$ in WM075, WM150. WM225, WM300 and WM600, respectively. Given the substantial distance between the top surface and the wall, we hypothesize that the effect of the top boundary condition on boundary layer development is minimal. 
At the top of the computational domain, in the WM-series, we apply the following boundary conditions: $v=v_{\mbox{blasius}}$, $\partial u/\partial y=\partial v/\partial x$, $\partial w/\partial y=\partial v/\partial z$.  \\

\begin{figure}
    \centering
    \includegraphics[width=0.8\linewidth]{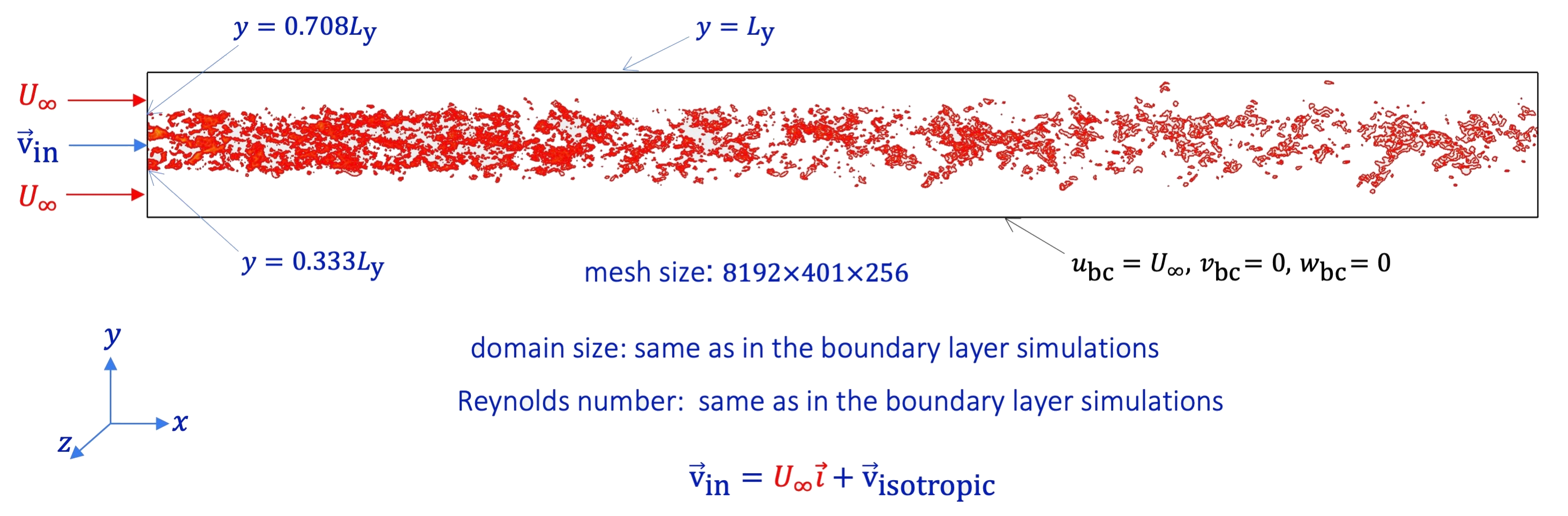}
    \caption{Illustration of spatially-developing isotropic turbulence DNS design using contours of streamwise velocity $u$ over a random $xy$-plane where blue and red color indicate low and higher values, respectively.  }
    \label{fig:sdit_dns_design}
\end{figure}

\noindent
For the SDIT-series DNSs, the domain size, inflow turbulence generation, and numerical method are the same as those in the WM-series (see Fig.~\ref{fig:sdit_dns_design}). The inflow turbulence is positioned in the central region of the inlet. The grid resolution (provided in Fig.~\ref{fig:sdit_dns_design}) is coarser than that in the WM series, as the absence of the wall reduces the demand for resolving the boundary-layer turbulence.

\section{DNS benchmarks for transition prediction}
\subsection{Data accessiblity}
Here, we present the processed skin friction, shape factor, dissipation and evolution of integral and Taylor length scales from our DNSs. For the sake of brevity, other quantities such as boundary-layer thickness ($\delta$), displacement thickness ($\delta^{*}$), wall-pressure fluctuation ($p_{w,rms}^{'}$), wall-shear stress fluctuation ($\tau_{w,rms}^{'}$) are not presented. However, the streamwise growth of these data with the streamwise ($Re_x$),  momentum thickness ($Re_\theta$) or the friction-based ($Re_\tau$) based Reynolds numbers are accessible from the \href{https://ctr.stanford.edu/about-center-turbulence-research/research-data}{Center for Turbulence Research website}. Additionally, the database contains the wall-normal variations of the mean velocity ($\overline{u}$) and of turbulent stresses ($u_{rms}^{'}$, $v_{rms}^{'}$, $w_{rms}^{'}$, $\overline{u^{'}v^{'}}$, $p_{rms}^{'}$, where overbar indicates averaging),
and total shear ($\tau$) in outer units ($y/\delta$) and viscous units ($y^{+}$) at selected streamwise stations covering the early, late and post-transition stages. The statistics were sampled on the fly at every time step over two convective flow-through times across the domain. The reader is referred to \citet{wu2017transitional,gonzalez2024benchmarks} for a discussion on the accuracy of the collected statistics. \\


\begin{figure}
    \centering
    \includegraphics[width=0.80\linewidth]{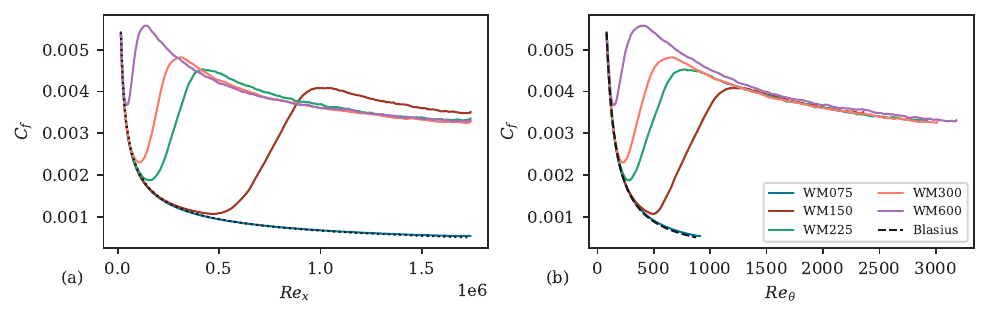}
    \caption{Skin-friction coefficient as a function of (a) streamwise Reynolds number and (b) momentum-thickness Reynolds number in the five boundary layer DNS cases.}
    \label{fig:cf_Rex}
\end{figure}

\subsection{Skin-friction and shape factor}
\begin{figure}
    \centering
    \includegraphics[width=0.5\linewidth]{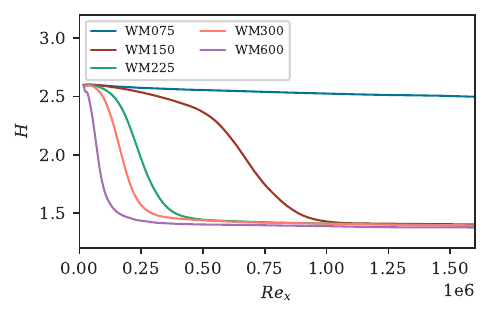}
    \caption{Boundary-layer shape factor as a function of momentum-thickness Reynolds number in the five boundary layer DNS cases. }
    \label{fig:h_Re_x}
\end{figure}

Fig.~\ref{fig:cf_Rex}(a) presents the skin friction, $C_{f}$ as a function of $Re_{x}$. All five DNSs show an extended domain range wherein the skin friction agrees well with the Blasius laminar flow solution. The departure from the Blasius solution is only visible slightly upstream of the minimum $C_{f}$ location (a qualitative marker of early transition). The case WM075 does not complete the transition process across the present computational domain. The distribution of $C_{f}$ against $Re_{\theta}$ is shown in Fig.~\ref{fig:cf_Rex}(b). Slightly downstream of the peak plateau, $C_f$ of WM225 flow collapses onto the WM300 profile. Similarly, the WM150 profile also collapses onto those of WM225 and WM300 after the peak, suggesting the downstream boundary layers after transition are turbulent with only minimal lingering transitional effects. The WM600 profile also nearly collapses with the other profiles, but the minor differences suggest that further higher freestream turbulence may cause substantial disturbances to the viscous sublayer of the turbulent boundary layer underneath (also see the boundary-layer shape factor $H$ of WM600 in Fig.~\ref{fig:h_Re_x}. The precipitous drop in the shape factor near the inlet for the WM600 case potentially implies a non-negligible disturbance on the boundary layer by the high freestream turbulence).


\subsection{Length scales}

Next, we present length scale information for the boundary layer DNSs. The rate of viscous dissipation $\varepsilon$ appears in the definitions of Kolmogorov length scale $\eta=(\nu^{3}/\varepsilon)^{1/4}$ and the large-eddy scale $L=\mbox{TKE}^{3/2}/\varepsilon$, $\mbox{TKE}=(u_{rms}^{'2}+v_{rms}^{'2}+w_{rms}^{'2})/2$.  The dissipation rate, $\varepsilon$, is also used in several transition and turbulence models \citep{launder1983numerical,shih1995realizable}. \\ 

\begin{figure}
    \centering
    \includegraphics[width=0.80\textwidth]{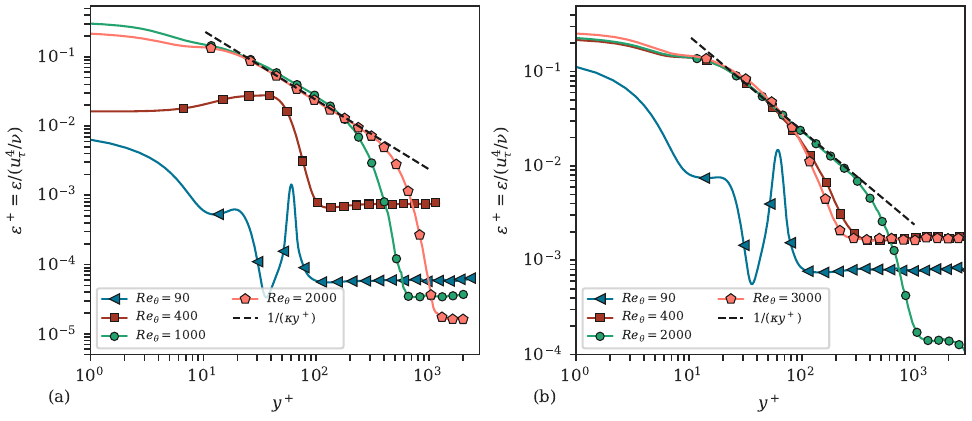}
    \caption{Rate of viscous dissipation $\varepsilon^{+}$ profiles before, during, and after transition in (a) WM150 and (b) WM600. }
    \label{fig:epsilon_WM150}
\end{figure}

\noindent
Fig.~\ref{fig:epsilon_WM150} presents wall-normal profiles of $\varepsilon^{+}=\varepsilon/(u_{\tau}^{4}/\nu)$ at four selected streamwise stations in WM150 and WM600. A narrow peak is observed at $y^{+}\approx 60$, corresponding to $y\approx 1.9\delta$, at the $Re_{\theta}=90$ station (which is slightly downstream of the inlet $Re_{\theta}=80$ station). We believe this may be associated with the inflow boundary setup: $\mathrm{FSTI}_{in}$ is only imposed for $y>2\delta$ (see Fig.~\ref{fig:inlet_design_dns}). For both WM150 and WM600 cases (in Fig.~\ref{fig:epsilon_WM150}), at stations downstream of the transition region, the dissipation profiles collapse well with the theoretical equation $\varepsilon^{+}=1/(\kappa y^{+})$ where $\kappa=0.421$ \citep{mckeon2004further} which can be derived by assuming constant shear stress layer, logarithmic velocity profile, and the equilibrium between the production of $\mbox{TKE}$ and $\varepsilon$ (also see \citet{tennekes1972first}). \\

\begin{figure}
    \centering
    \includegraphics[width=0.80\textwidth]{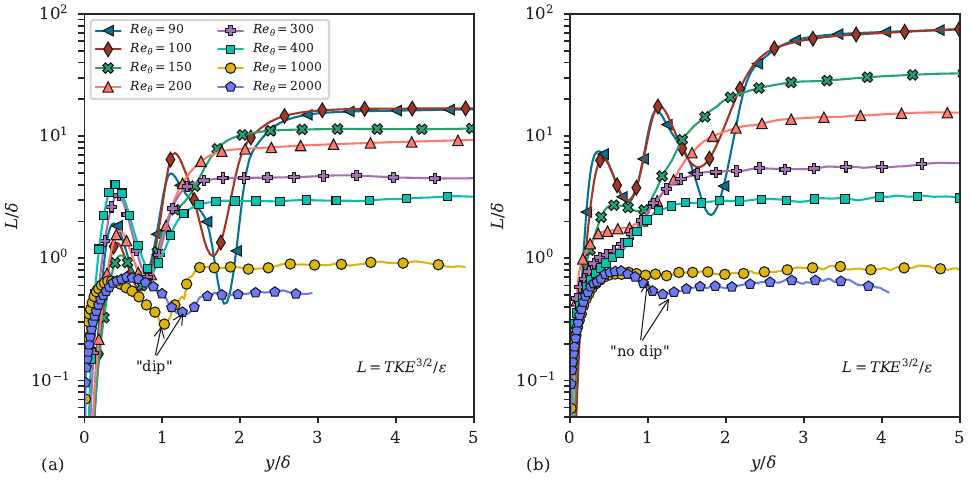}
    \caption{Profiles of large-eddy scale $L/\delta$ before, during and after transition in (a) WM150 and (b) WM600. }
    \label{fig:L_WM150}
    \end{figure}

\noindent
Fig.~\ref{fig:L_WM150}(a) presents the wall-normal profiles of $L/\delta$ at eight selected streamwise stations in WM150 where $\delta$ is the local boundary layer thickness. The nearly constant values of $L/\delta$ for $y>2.5\delta$ represent the freestream values. The sharp descent and rapid rise in the profile of $Re_{\theta}=90$ near $1.8\delta<y<2\delta$ is related to the peak of $\varepsilon$ near $y\approx 1.9\delta$ (see Fig.~\ref{fig:epsilon_WM150}). 
As $Re_\theta$ increases (moving downstream), there is a successive reduction in the large-eddy length scale relative to the local boundary-layer thickness. 
For WM150 case, in Fig.~\ref{fig:L_WM150}(a), there is a noticeable “dip” near the boundary-layer edge in the profiles of $Re_{\theta}=1000$ and $2000$. We believe that the minima of such a “dip” corresponds to the boundary-layer turbulence and freestream turbulence interface (BTFTI) separating the two types of turbulent motions. Instantaneous BTFTIs of WM300 were studied in \citet{wu2019boundary} using the probability density functions of passive scalar and vorticity.  Fig.~\ref{fig:L_WM150} provides a hint on a new indicator for locating the BTFTI. The absence of a “dip” in the WM600 profiles is consistent with the physical mechanisms that strong perturbations disturb the boundary layer underneath the FST. This interaction obscures the distinctions between the two turbulent regions, supporting our conjecture that beyond $\mathrm{FSTI}_{in} \approx 6\%$, the boundary layer may get significantly affected by the freestream turbulence. \\


\begin{figure}
    \centering
    \includegraphics[width=0.80\textwidth]{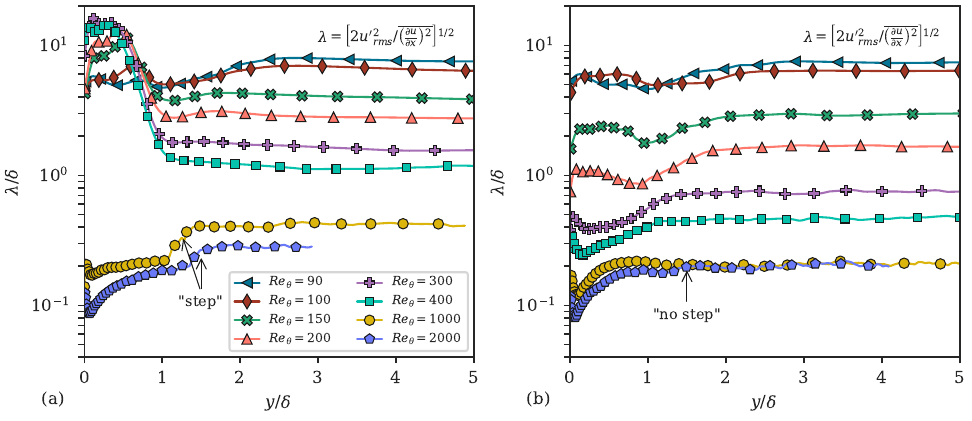}
    \caption{Profiles of Taylor-microscale $\lambda/\delta$ before, during and after transition in (a) WM150 and (b) WM600. }
    \label{fig:lambda_WM150}
\end{figure}

\noindent
Fig.~\ref{fig:lambda_WM150} presents the wall-normal profiles of the normalized Taylor-microscale ($\lambda/\delta$) at eight selected streamwise stations for WM150 and WM600. Here $\lambda^{2}=2u_{rms}^{'2}/\overline{(\partial u/\partial x)^2}$. The freestream values of $\lambda/\delta$ decrease along the downstream direction, indicating that the freestream Taylor-microscale grows more slowly than the boundary layer. For the WM150 case, there is a noticeable “step” near the boundary-layer edge bridging the near-wall flow and the FST in both the turbulent region profiles of $Re_{\theta}=1000$ and 2000. We also consider these statistical markers of the BTFTI. Similarly, there is a lack of a “step” in Fig.~\ref{fig:lambda_WM150}(b) for the WM600 case at $Re_{\theta}=1000$ and 2000 stations.


\section{Decay of freestream turbulence}

In this section, we compare the streamwise developments of the turbulence intensities, viscous dissipation $\varepsilon$, large-eddy scale $L$, and Taylor-microscale $\lambda$ for spatially developing isotropic turbulence and compare it to the corresponding freestream flow of a transitional boundary layer. \\

\begin{figure}
    \centering
    \includegraphics[width=0.80\linewidth]{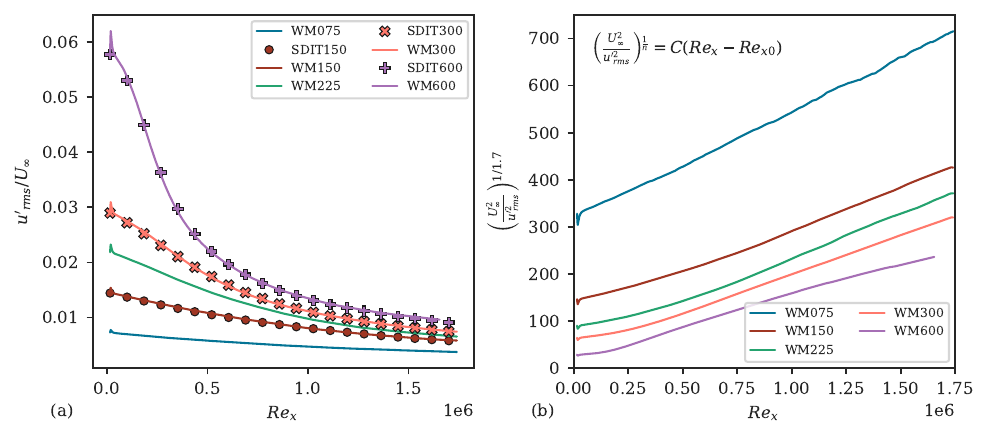}
    \caption{(a) Decay of freestream turbulence intensity in the WM-series and the SDIT-series cases and (b) fitting of the freestream turbulence intensity decay into a power-law with exponent $1.7$ in the WM-series. }
    \label{fig:fst_decay_intensity}
\end{figure}

\noindent
Fig.~\ref{fig:fst_decay_intensity} presents the decay of 
$u_{rms}^{'}/U_{\infty}$ as a function of $Re_{x}$. For the WM-series, the $u_{rms}^{'}$ at each $Re_{x}$ station is obtained by averaging the freestream values over the wall-normal range $500\theta_{in}<y<800\theta_{in}$ (see Fig.~\ref{fig:inlet_design_dns}). For the SDIT-series, the $u_{rms}^{'}$ is obtained by averaging the values over the wall-normal range $0.4L_{y}<y<0.6L_{y}$ (see Fig.~\ref{fig:sdit_dns_design}). The decay of turbulence intensity in the WM-series is nearly identical to the SDIT-series counterparts, implying that the presence of the wall (and the associated boundary layer) has a minimal effect on the rate of the FST decay. In Fig.~\ref{fig:fst_decay_intensity}(b), assuming a power-decay, $U_{\infty}^{2}/u_{rms}^{'2}=a(Re_{x}-Re_{x_{0}})^{n}$, we plot $Re_{x}$ versus $(U_{\infty}^{2}/u_{rms}^{'2})^{1/n}$ to identify the exponent $n$ where $x_{0}$ is an unknown virtual origin of the grid turbulence. For the WM075 flow, the entire streamwise growth is nearly linear. For the higher FST level flows, the linear region only appears for $Re_{x}>1.0\times 10^{6}$. The value of $n=1.7$ is also consistent with the experiments of \citet{ling1972decay} for isotropic turbulence. \\ 

\begin{figure}
    \centering
    \includegraphics[width=0.80\linewidth]{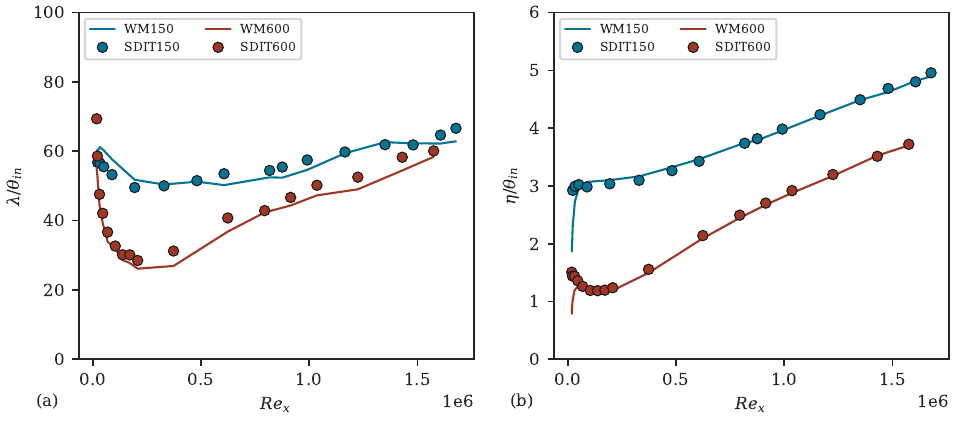}
    \caption{Comparison of the freestream turbulence length scale variations in the WM-series and those in the SDIT-series. In the WM-series, data points are evaluated at $y=2\delta(x)$. In the SDIT-series, data points are averaged over the wall-normal range $0.4L_{y}<y<0.6L_{y}$. (a) Taylor-microscale; (b) Kolmogorov length scale. }
    \label{fig:fst_decay_scale}
\end{figure}

\noindent
Next, we assess the development of the turbulence length scales in the freestream region of the WM-series and SDIT-series flows. Fig.~\ref{fig:fst_decay_scale}a compares the variations of $\lambda/\theta_{in}$ with $Re_{x}$ between WM150 and SDIT150, and between WM600 and SDIT600. The freestream $\lambda$ value is obtained from $y=2\delta(x)$ location for the WM series. The Taylor-microscale for the WM-series flows develops similarly to their corresponding SDIT flows. Similarly, in Fig.~\ref{fig:fst_decay_scale}b, we find that the Kolmogorov scales between the WM-series and SDIT-series flows also evolve similarly (the freestream Kolmogorov length scale value is also obtained from $y=2\delta(x)$ for WM-series flows).  

\section{Blasius layer breakdown mechanisms at intermediate FSTI}

Within the context of bypass transition in the narrow sense, the mechanism of breakdown of a perturbed Blasius layer is synonymous with the mechanism of the inception of the turbulent spot at the smallest $Re_{x}$, which is generally quasi-cyclic. We believe that the inception of additional turbulent spots within a cycle further downstream of the existing infant turbulent spot (closest to the inlet) may facilitate the growth of the turbulent region but may not be directly responsible for the Blasius layer breakdown. \\

\noindent
In Subsection 1.2, we reviewed previous DNS work concerning the role of streak primary instability and secondary instability leading to the breakdown of the Blasius layer in boundary-layer bypass transition in the narrow sense. For instance, \citet{jacobs2001simulations} found streak primary instability relevant to the breakdown, \citet{brandt2004transition} and \citet{alarcon2024role} advocated the roles of streak primary and secondary instabilities. On the other hand, \citet{ovchinnikov2008numerical} and \citet{wu2017transitional}
provided evidence that infant turbulent spots form upstream of streak primary instability, hence streaks are not responsible for the Blasius layer breakdown in their particular $\mathrm{FSTI}_{in}$ cases. \citet{wu2023new} reconciled the differences and conjectured that streak primary and secondary instabilities are important for $\mathrm{FSTI}_{in} \leq 2\%$, whereas the self-amplifying process of oblique vortex filaments interacting with a $\Delta$-shaped low-speed patch underneath are important for FSTI greater than $2.5\%$. \\

\begin{figure}
    \centering
    \includegraphics[width=0.4\linewidth]{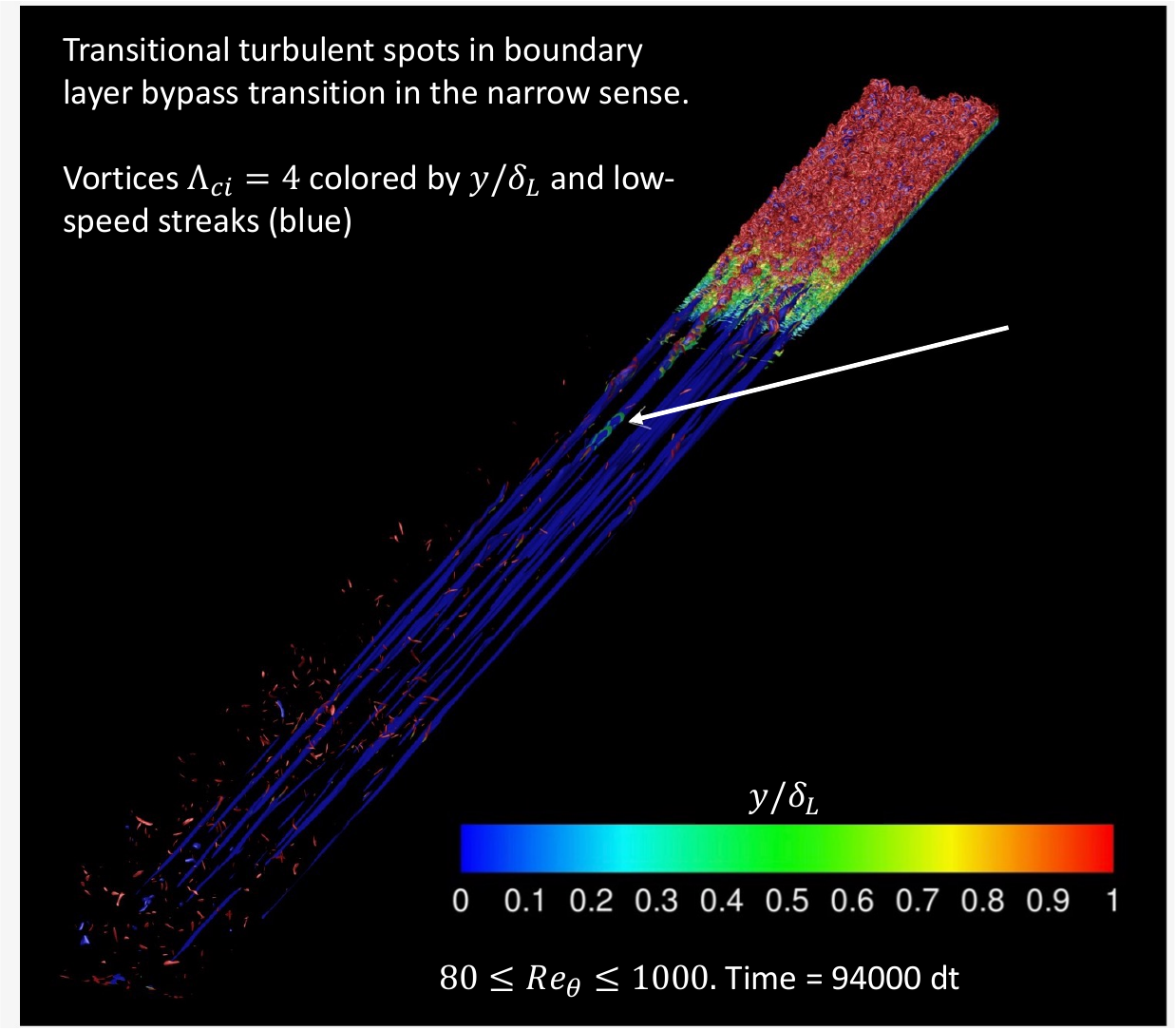}
    \caption{Visualization of the inception of an infant turbulent spot in WM150 due to streak primary and secondary instabilities. Streaks are revealed using blue-colored iso-surfaces of $u^{'}=-0.1U_{\infty}$. Vortices are revealed using iso-surfaces of swirling strength colored by $y/\delta(x)$. }
    \label{fig:infant_spot_view_wm150}
\end{figure}

\noindent
For the WM150 flow, Fig.~\ref{fig:infant_spot_view_wm150} presents isosurfaces of swirling strength, as defined in \citet{adrian2007hairpin}, colored by $y/\delta$, and blue-colored isosurfaces of low-speed fluid $u^{'}=-0.1U_{\infty}$. The long streaks indicate primary instability, and the highlighted two symmetric vortex filaments wrapping around one streak correspond to the ‘varicose mode’ streak secondary instability. The structure subsequently evolves into an infant turbulent spot, with no other turbulent spots in this cycle at a smaller $Re_{x}$. Thus, this particular spot may be directly responsible for the breakdown of the Blasius layer within this cycle (movies for this flow can be found in \citet{wu2023new}). \\

\begin{figure}
    \centering
    \includegraphics[width=0.4\linewidth]{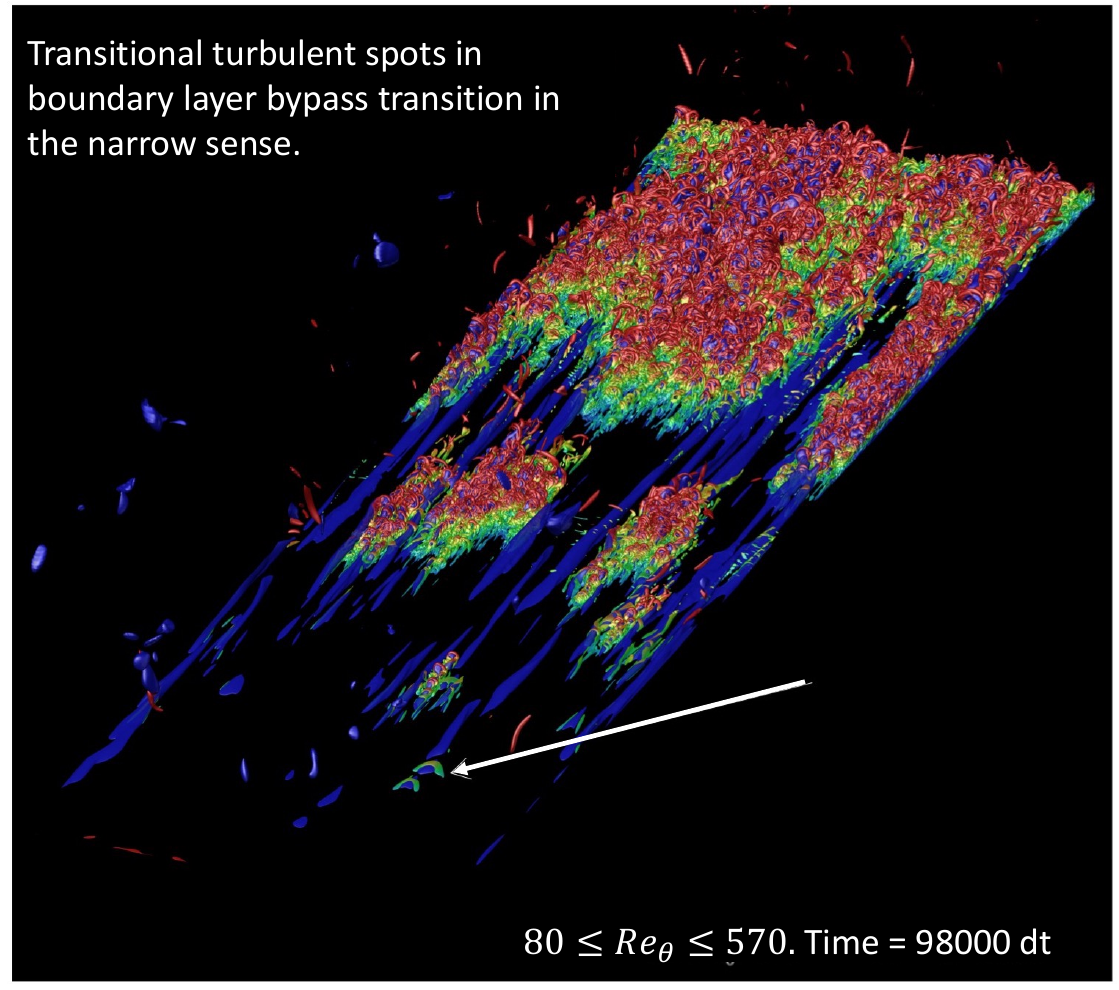}
    \caption{Visualization of the inception of an infant turbulent spot in WM300. The process is not due to streak instabilities. Streaks are revealed using blue-colored iso-surfaces of $u^{'}=-0.1U_{\infty}$. Vortices are revealed using iso-surfaces of swirling strength colored by $y/\delta(x)$. }
    \label{fig:infant_spot_view_wm300}
\end{figure}

\noindent
Similarly, for WM300 flow, Fig.~\ref{fig:infant_spot_view_wm300} presents a set of isosurfaces of swirling strength, colored by $y/\delta$, and blue-colored isosurfaces of low-speed fluid $u=-0.1U_{\infty}$. As highlighted by the long arrow, at the instant $t=98000\Delta t$, there are two oblique vortex filaments, each interacting with a $\Delta$-shaped low-speed patch underneath. Subsequently, the structures evolve into an infant spot with no other upstream spots in this cycle. Long streaks appear only downstream of this infant turbulent spot, in agreement with the observation of \citet{ovchinnikov2008numerical} at $\mathrm{FSTI}_{in} > 6\%$. Streak instabilities are not found to cause the Blasius layer breakdown in this case; however, they may facilitate the spreading of turbulent regions (movies for this case can also be found in \citet{wu2023new}). \\

\begin{figure}
    \centering
    \includegraphics[width=0.8\linewidth]{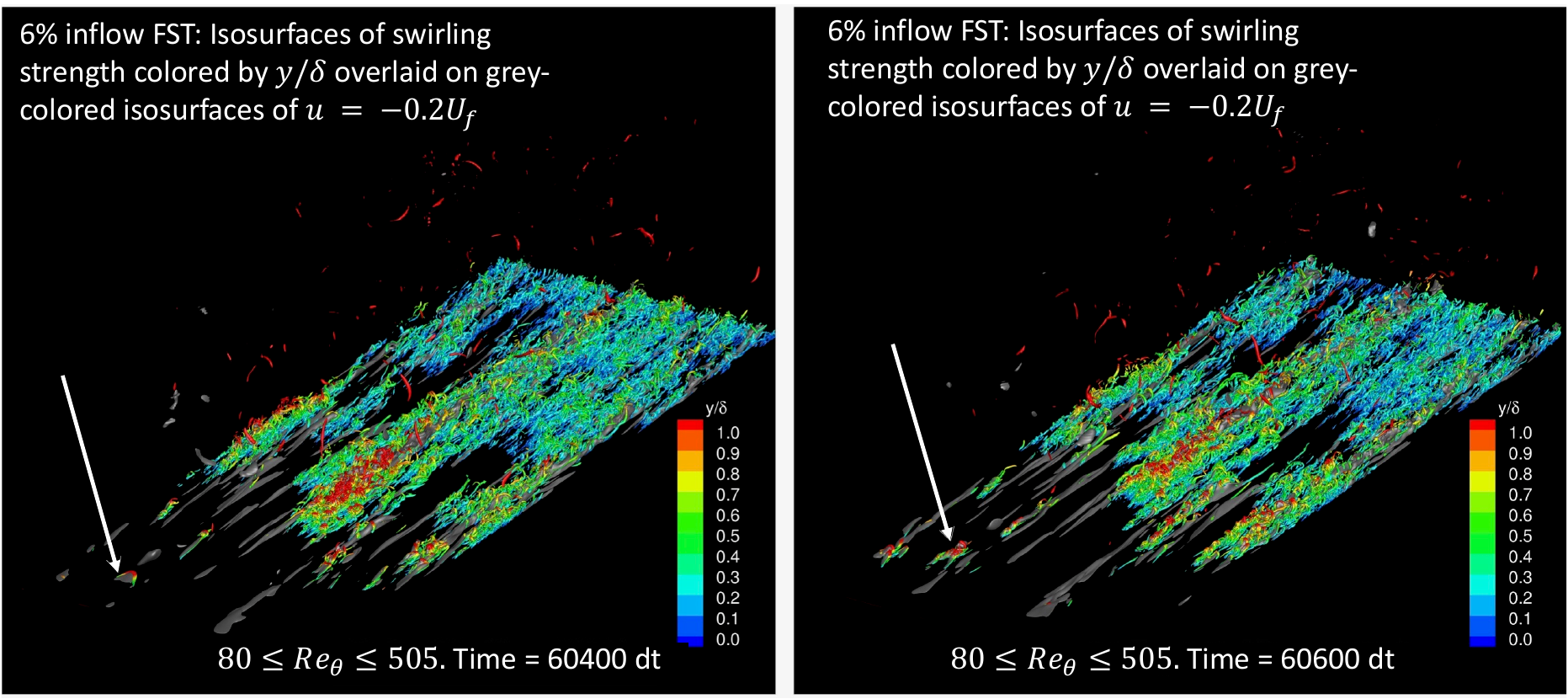}
    \caption{Visualization of the inception of an infant turbulent spot in WM600. The process is not due to streak instabilities. 
    Streaks are revealed using grey-colored iso-surfaces of $u^{'}=-0.2U_{\infty}$. Vortices are revealed using iso-surfaces of swirling strength colored by $y/\delta(x)$. }
    \label{fig:infant_spot_view_wm600_a}
\end{figure}

\noindent
Finally, for WM600 flow,  Fig.~\ref{fig:infant_spot_view_wm600_a} presents the isosurfaces of swirling strength, also colored by $y/\delta$, and grey-colored isosurfaces of low-speed fluid $u=-0.2U_{\infty}$. At $6\%$ FSTI, the transitional region is more chaotic than in other cases. As highlighted by a long arrow, at the instant $t=60400\Delta t$, an oblique vortex filament interacts with a $\Delta$-shaped low-speed patch underneath with no other spots further upstream. Subsequently, at $t=60600\Delta t$, the structures evolve into an infant spot.
Streaks appear downstream of this infant turbulent spot. Streak instabilities are not crucial in the inception of infant turbulent spots; however, they may aid in the growth of the turbulent region: when nearby existing turbulent spots interact with a streak, the streak may oscillate and break down. 
Supporting supplemental movies (WM600-Vortices, WM600-Streaks, and WM600-Streaks-and-Vortices) provide further evidence that at $\mathrm{FSTI}_{in} = 6.0\%$, the breakdown is governed by the self-amplifying process of oblique vortex filaments interacting with a $\Delta$-shaped low-speed patch underneath.  \\

\begin{figure}
    \centering
    \includegraphics[width=0.7\linewidth]{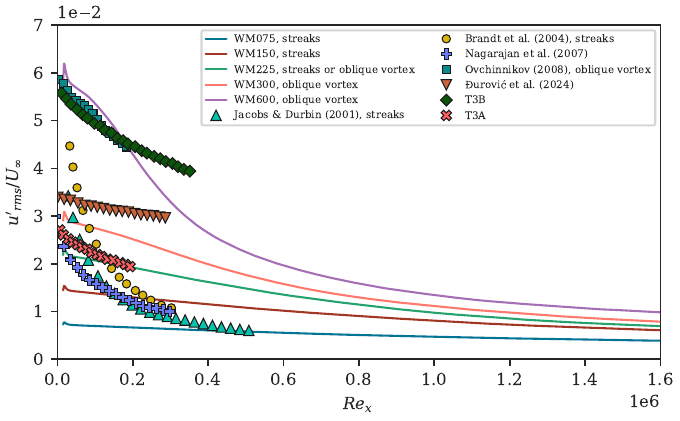}
    \caption{Decay of freestream turbulence intensity in the present and previous studies on bypass transition in the narrow sense. }
    \label{fig:fst_decay_compare}
\end{figure}

\noindent
\citet{wu2023new} conjectured that the mechanism of self-amplifying oblique-vortex interaction with an underneath $\Delta$-shaped low-speed patch is likely to be relevant at high inlet FST levels ranging from $2.5\%$ to $6\%$. Observations in support of this conjecture include the WM300 and WM600 results and those of \citet{ovchinnikov2008numerical}. \citet{wu2023new} also conjectured that it is quite likely that the streak-centered turbulent spot inception mechanism is relevant at $\mathrm{FSTI}_{in}$ values ranging from $0.5\%$ to $2\%$. The present WM150 results fall into this category. However, previous studies of \citet{jacobs2001simulations,brandt2004transition,alarcon2024role} reported that this streak-centered mechanism is still relevant even at their higher FSTI values of $3.5\%$, $3.45\%$ and $4.7\%$, respectively. To reconcile this apparent contradiction, we refer the reader to Fig.~\ref{fig:fst_decay_compare}.
The FST introduced from the inlet in these simulations rapidly (for instance, compared to experiments of \citet{roach1990influence}). The freestream turbulence intensity in \citet{brandt2004transition} is approximately $4.7\%$ at the inlet but drops down to $1.2\%$ by the station $Re_{x}=2.5\times 10^{5}$. Similarly, in \citet{jacobs2001simulations}, the FST decays from $3.5\%$ at the inlet to $0.9\%$ by the station $Re_{x}=2.5\times 10^{5}$. In contrast, the present DNS profiles show a slower decay than the existing DNS profiles. For example, in WM300, at $Re_{x}=2.5\times 10^{5}$, the $\mathrm{FSTI}_{in} = 3.0\%$ case only decays to $2.5\%$. Thus, we conjecture that the bypass transition reported in these previous DNS might be occurring in a weaker FST environment than the reported corresponding $\mathrm{FSTI}_{in}$, which would aid in reconciling the previous apparent contradiction. We briefly remark that, although, for \citet{alarcon2024role} (who utilize the DNs of \citet{djurovic2024direct}), at the station, $Re_x = 2.5 \times 10^5$, the freestream turbulence is higher than 3\%, their $C_f$ departs from the Blasius solution in the laminar region early, which is unexpected. \\


\noindent
Admitting the conjecture that the streak-centered turbulent spot inception mechanism is relevant at mild inlet FST levels ranging from $0.5\%$ to $2\%$, and the mechanism of self-amplifying oblique-vortex interaction with an underneath $\Delta$-shaped low-speed patch is relevant at stronger inlet FST levels ranging from $2.5\%$ to $6\%$, we assess the intermediate zone between FSTI $2.0\%$ and $2.5\%$. For instance, finding evidence of the co-existence of the two mechanisms in one flow at an intermediate FSTI would lend more support to our conjectures. \\

\begin{figure}
    \centering
    \includegraphics[width=0.8\linewidth]{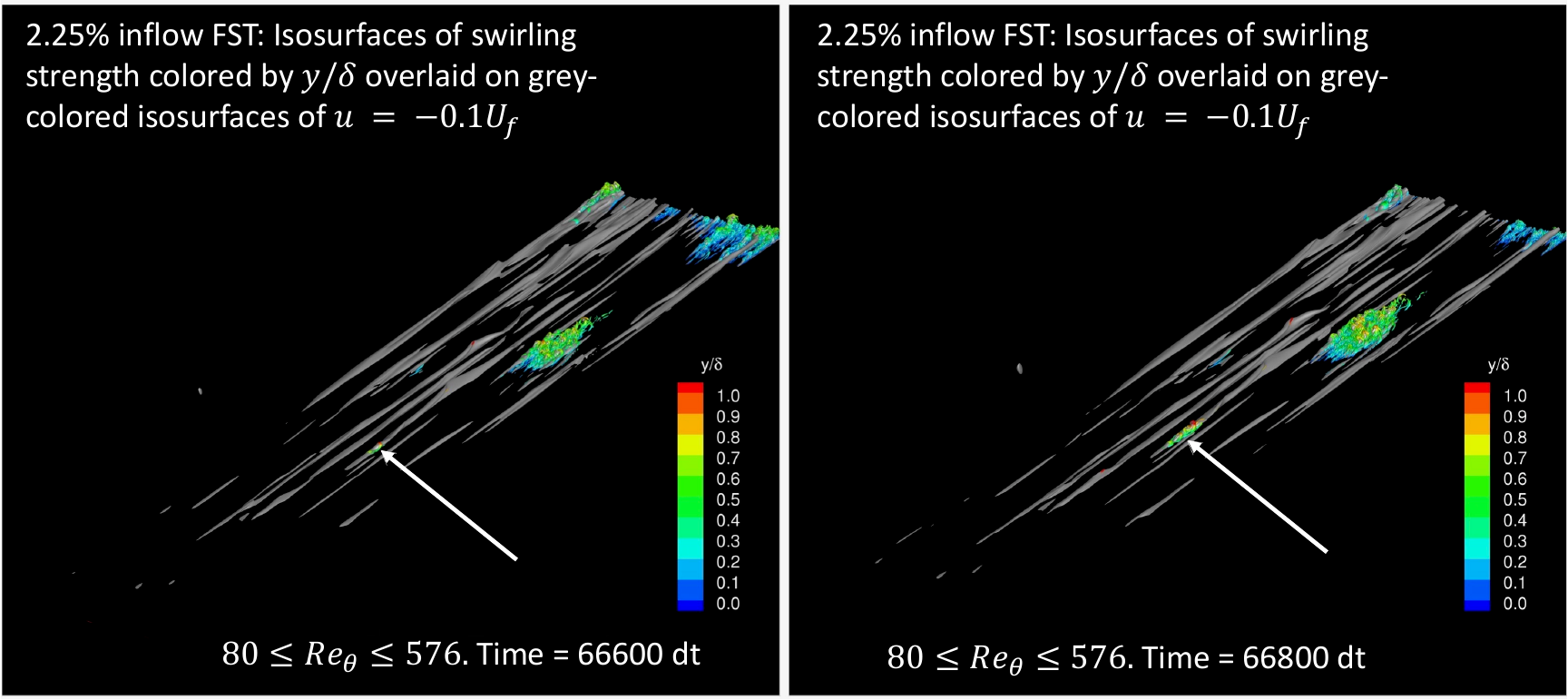}
    \caption{Visualization of the inception of an infant turbulent spot in WM225 due to streak primary and secondary instabilities. Streaks are revealed using grey-colored iso-surfaces of $u^{'}=-0.1U_{\infty}$. Vortices are revealed using iso-surfaces of swirling strength colored by $y/\delta(x)$. }
    \label{fig:infant_spot_view_wm225_a}
\end{figure}

\begin{figure}
    \centering
    \includegraphics[width=0.8\linewidth]{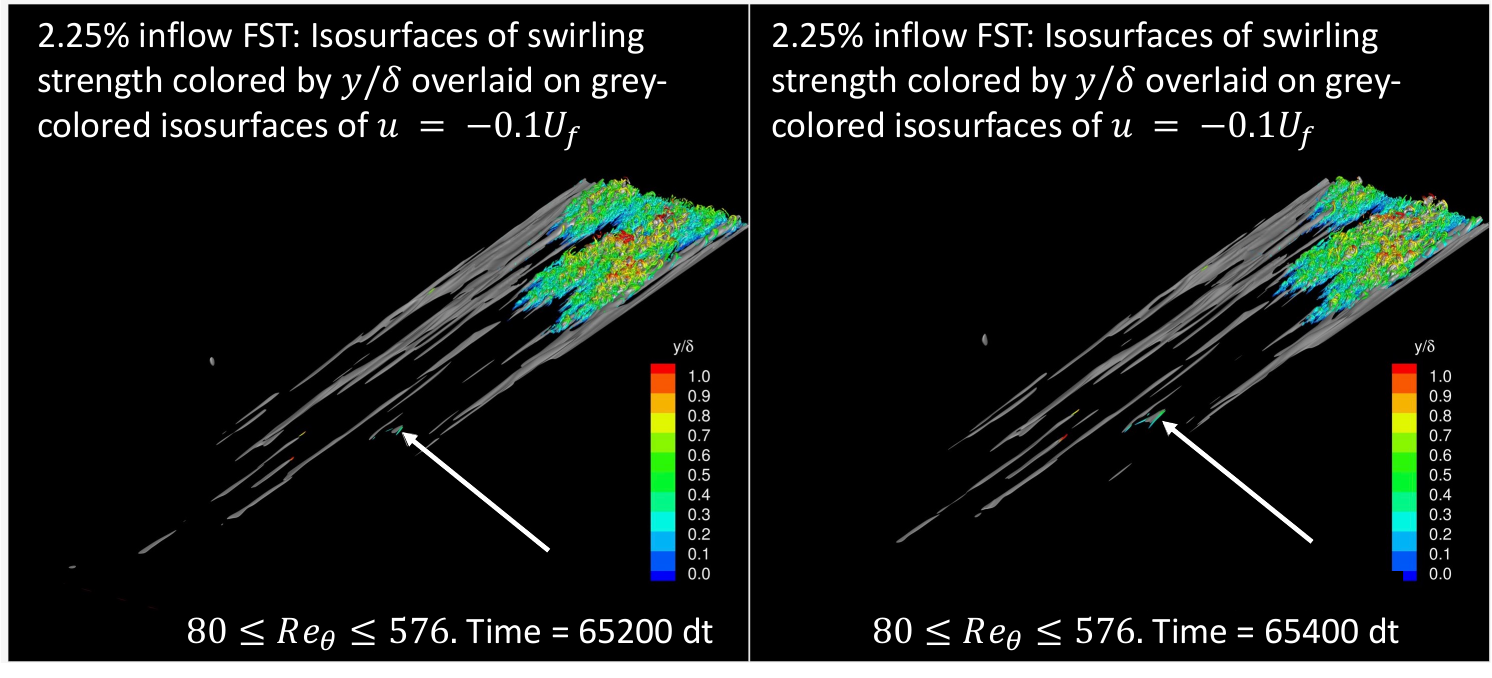}
    \caption{Visualization of the inception of an infant turbulent spot in WM225 not due to streak instabilities. Streaks are revealed using grey-colored iso-surfaces of $u^{'}=-0.1U_{\infty}$. Vortices are revealed using iso-surfaces of swirling strength colored by $y/\delta(x)$. }
    \label{fig:infant_spot_view_wm225_b}
\end{figure}

\noindent
In this spirit, for WM225 flow, Fig.~\ref{fig:infant_spot_view_wm225_a} presents, at two consecutive instants, iso-surfaces of the vortex identifier swirling strength, colored by $y/\delta$, overlaid on the grey-colored iso-surfaces of low-speed fluid $u/U_{\infty}=-0.1$. Regular and long low-speed streaks are abundant. In the figure, at $t=66600\Delta t$, symmetric boundary-layer vortex filaments (small hairpins) partially wrap the tail portion of one such streak. This results in the formation of an infant turbulent spot subsequently at $t=66800\Delta t$ with no other spots further upstream in this cycle. The Blasius layer breakdown in this cycle is due to streak primary and secondary instabilities.
For the same flow of WM225, at another two consecutive instants, Fig.~\ref{fig:infant_spot_view_wm225_b}  reveals the other breakdown mechanism. At $t=65200\Delta t$, a pair of oblique vortex filaments flank a $\Delta$-shaped low-speed patch underneath (in $y<0.7\delta$). Note there is no turbulent spot upstream of this oblique vortex, and this structure is at the smallest $Re_{x}$ in this cycle. Growth and amplification of the oblique vortex and the $\Delta$-shaped low-speed patch can be seen at $t=65400\Delta t$. The Blasius layer breakdown in this cycle is due to the oblique vortex mechanism, not streaks (full dynamic views of the two breakdown paths in the WM225 flow can be found in Supplemental Movies: 
WM225-Vortices, WM225-Streaks, and WM225-Streaks-and-Vortices. These three movies cover the transition stage and show the development history of boundary-layer vortices, low-speed structures, and vortices overlaid onto the low-speed structures, respectively). \\

\noindent
Thus, we believe that there is sufficient evidence demonstrating the co-existence of the following two breakdown mechanisms in a given flow of the boundary-layer bypass transition at $\mathrm{FSTI}_{in} = 2.25\%$: the mechanism of long low-speed streak primary instability followed by the ‘sinuous mode’ and ‘varicose mode’ secondary instability and the self-amplifying process of oblique vortex filaments interacting with a $\Delta$-shaped low-speed patch underneath. Here, only the event of the inception of the one infant turbulent spot closest to the inlet in a cycle is considered instrumental in the laminar layer breakdown. Thus, at any instant, there can, at most, be only one breakdown mechanism at play. In this work, co-existence refers to both mechanisms being found in the same flow at different time instants. 
 
\section{Concluding Remarks}

This work provides a new database of comprehensive direct numerical simulation benchmarks of bypass transition. Detailed statistics and descriptions of the length-scales and dissipation are also provided for freestream turbulence intensity (FSTI) levels of $0.75\%$, $1.5\%$, $2.25\%$, $3.0\%$, $6.0\%$. We show that the presence of a wall (and the growing boundary layer) does not significantly affect the decay rates or the turbulence length scales (Taylor and Kolmogorov) in the freestream flow compared to its spatially evolving homogeneous flow counterpart. Further, this study reveals important insights into the transition process across various FSTI levels. Specifically, at an intermediate FSTI equal to 2.25\%, we provide evidence for the co-existence of two distinct breakdown mechanisms: the low-speed streak primary instability known to exist at low FSTI and the self-amplifying oblique vortex filaments known to exist at high FSTI. We also provide evidence that a reasonable upper limit for bypass transition in the narrow sense may be FSTI $\approx 6\%$. \\

\section*{Acknowledgements}

This work was pursued during the 2024 Center for Turbulence Research Summer Program. X.W. is supported by the NSERC Discovery Grant and Digital Alliance Canada. C.A.G. and R.A. were supported by the NASA Transformational Tools and Technologies Program (grant \#80NSSC20M0201). R.A. also acknowledges support from Boeing Research and Technology (grant \#2024-UI-PA-100) and the Franklin P. and Caroline M. Johnson Fellowship at Stanford University. The authors gratefully acknowledge helpful discussions with Prof. Parviz Moin. 


\end{document}